\documentclass[12pt]{article}
\usepackage{graphicx}
\usepackage{color}
\usepackage{cite}

\def\upstrut{\vrule height 2.5ex depth 0.0ex width 0pt}

\def\mystrut{\vrule height 3.5ex depth 2.2ex width 0pt}
\def \beq{\begin{equation}}
\def \eeq{\end{equation}}
\def\eqref#1{(\ref{#1})}
\def\bea{\begin{eqnarray}}
\def\eea{\end{eqnarray}}
\def\jpsi{J/\psi}

\def\URLtilde{\lower0.2em\hbox{$\tilde{\phantom{a}}$}}
\def\mycomm#1{\hfill\break\strut\kern-3em{\color{red}\tt ====> #1
\color{black}}\hfill\break}

\newcount\timecount
\newcount\hours \newcount\minutes  \newcount\temp \newcount\pmhours
\hours = \time
\divide\hours by 60
\temp = \hours
\multiply\temp by 60
\minutes = \time
\advance\minutes by -\temp
\def\hour{\the\hours}
\def\minute{\ifnum\minutes<10 0\the\minutes
\else\the\minutes\fi}
\def\clock{
\ifnum\hours=0 12:\minute\ AM
\else\ifnum\hours<12 \hour:\minute\ AM
\else\ifnum\hours=12 12:\minute\ PM
\else\ifnum\hours>12
\pmhours=\hours
\advance\pmhours by -12
\the\pmhours:\minute\ PM
\fi
\fi
\fi
\fi
}

\def\monthname{\relax\ifcase\month 0/\or January\or February\or
March\or April\or May\or June\or July\or August\or September\or
October\or November\or December\else\number\month/\fi}

\def\bold#1{\setbox0=\hbox{$#1$}     \kern-.025em\copy0\kern-\wd0
\kern.05em\copy0\kern-\wd0
\kern-.025em\raise.0433em\box0 }

\begin{document}
\setcounter{footnote}{1}
\rightline{EFI 14-28}
\rightline{TAUP 2986/14}
\rightline{arXiv:1408.5877}
\vskip1.5cm

\centerline{\large \bf Baryons with two heavy quarks:}
\centerline{\large \bf Masses, production, decays, and detection}
\bigskip

\centerline{Marek Karliner$^a$\footnote{{\tt marek@proton.tau.ac.il}}
 and Jonathan L. Rosner$^b$\footnote{{\tt rosner@hep.uchicago.edu}}}
\medskip

\centerline{$^a$ {\it School of Physics and Astronomy}}
\centerline{\it Raymond and Beverly Sackler Faculty of Exact Sciences}
\centerline{\it Tel Aviv University, Tel Aviv 69978, Israel}
\medskip

\centerline{$^b$ {\it Enrico Fermi Institute and Department of Physics}}
\centerline{\it University of Chicago, 5620 S. Ellis Avenue, Chicago, IL
60637, USA}
\bigskip
\strut

\begin{center}
ABSTRACT
\end{center}
\begin{quote}
The large number of $B_c$ mesons observed by LHCb suggests a sizable
cross section for producing doubly-heavy baryons in the same experiment.
Motivated by this, we estimate masses of the doubly-heavy $J=1/2$ baryons
$\Xi_{cc}$, $\Xi_{bb}$, and $\Xi_{bc}$, and their $J=3/2$ hyperfine partners,
using a method which accurately predicts the masses of ground-state baryons
with a single heavy quark.  We obtain
$M(\Xi_{cc})  =   3627 \pm 12$ MeV,
$M(\Xi_{cc}^*)=   3690 \pm 12$ MeV,
$M(\Xi_{bb})  =  10162 \pm 12$ MeV,
$M(\Xi_{bb}^*)=  10184 \pm 12$ MeV,
$M(\Xi_{bc}) =    6914 \pm 13$ MeV,
$M(\Xi'_{bc}) =   6933 \pm 12$ MeV,
and
$M(\Xi_{bc}^*) =  6969 \pm 14$ MeV.
As a byproduct, we estimate the hyperfine splitting between $B_c^*$ and $B_c$
mesons to be $68 \pm 8$ MeV.  We discuss P-wave excitations, production
mechanisms, decay modes, lifetimes, and prospects for detection of the doubly
heavy baryons.
\end{quote}
\smallskip

\leftline{PACS codes: 14.20.Lq, 14.20.Mr, 12.40.Yx}
\bigskip


\section{Introduction \label{sec:intro}}
Some simple arguments based on the quark model have been shown to
accurately predict the
spectrum of baryons containing a single $b$ quark \cite{Karliner:2008sv,%
Karliner:2006ny}.  The question then arises:  Can such methods be applied to
systems with two or more heavy quarks?  So far the only experimental
evidence for such
states comes from the SELEX experiment, which has reported a state at 3520 MeV
containing two charm quarks and a down quark \cite{Mattson:2002vu,%
Ocherashvili:2004hi}, with a conference report of states at 3460 MeV and 3780
MeV containing two charm quarks and an up quark \cite{Engelfried:2007at}.
Despite several searches 
\cite{Ratti:2003ez,Chistov:2006zj,Aubert:2006qw,Aaij:2013voa,Kato:2013ynr}, 
no other experiment has confirmed this result. On the optimistic
side, one should notice that a large number of $B_c$ mesons has been seen both
by the Tevatron experiments \cite{Aaltonen:2007gv,Abazov:2008kv} and by LHCb 
\cite{Aaij:2012dd, Aaij:2014asa,Aaij:2014jxa,Aaij:2014bva,Aaij:2013cda,%
Aaij:2013vcx,Aaij:2013gia}.  From this one can infer \cite{Karliner:2013dqa} a
substantial cross section for simultaneous production of two pairs of heavy
quarks and their subsequent coalescence into a doubly-heavy hadron.

In this paper we estimate the mass of the lowest-lying $J=1/2$ $ccu$ or $ccd$
state, finding a value consistent with many other estimates lying well above
the SELEX results.  We estimate its branching fractions to various final states
and discuss the possibility of observing $bcu$, $bcd$, $bbu$, and $bbd$
ground-state baryons.  We also estimate the masses of the hyperfine ($J=3/2$)
partners of these states, comment briefly on P-wave excitations, and
discuss production, decays, and detection of these states.

In order to have a self-contained discussion, we review calculations based
on similar methods for baryons and mesons containing only $u,d,$ and $s$
quarks (Sec.\ \ref{sec:uds}) and those containing a single charmed quark
(Sec.\ \ref{sec:onec}) or a single bottom quark (Sec.\ \ref{sec:oneb}).  These
last two sections also include for completeness discussions of states with
both charm (or beauty) and strangeness.  Although we do not discuss $ccs$,
$bcs$, or $bbs$ states in the present paper, regarding their observation as
far in the future, we give enough information that their masses may be
readily calculated using the present methods.

In what follows we shall neglect the difference between the masses of $u$ and
$d$, referring to them collectively as $q$.  Masses of states with nonzero
isospin are taken to be isospin averages.  (Isospin splittings of doubly
heavy baryons are expected not to exceed several MeV \cite{Brodsky:2011zs,%
Borsanyi:2014jba}.) We calculate the masses of the
lowest-lying states of $ccq$ in Sec.\ \ref{sec:ccq}, $bbq$ in Sec.\
\ref{sec:bbq}, and $bcq$ in Sec.\ \ref{sec:bcq}, commenting briefly on
P-wave excitations in Sec.\ \ref{sec:pw}.  Likely decay modes are noted
in Sec.\ \ref{sec:br}, some suggestions for observing the states are made in
Sec.\ \ref{sec:prod}, while Sec.\ \ref{sec:concl} concludes.

\section{States containing only $u$, $d$, and $s$ quarks \label{sec:uds}}

\subsection{Baryons}

The following contributions suffice to describe the ground-state baryons
containing $u,d,s$ \cite{DeRujula:1975ge,Gasiorowicz:1981jz}.
\begin{itemize}

\item The effective masses of the $u$, $d$, and $s$ quarks

\item Their mutual hyperfine interactions

\end{itemize}
\noindent
(With the addition of heavy-quark masses, these methods were already used
in Refs.\cite{DeRujula:1975ge} and \cite{bj:heavy_baryon_masses}
to estimate masses of baryons with two heavy quarks.)

\begin{table}[h]
\caption{Quark model description of ground-state baryons containing $u,d,s$.
Here we take $m^b_u = m^b_d \equiv m^b_q = 363$ MeV, $m^b_s = 538$ MeV, and
hyperfine interaction term $a/(m^b_q)^2 = 50$ MeV.
\label{tab:lqb}}
\begin{center}
\begin{tabular}{c c c c} \hline \hline
State (mass     & Spin &   Expression for mass   & Predicted  \\
in MeV)         &     & \cite{Gasiorowicz:1981jz} & mass (MeV) \\ \hline
$N(939)$        & 1/2 & $3m^b_q - 3a/(m^b_q)^2$       &    939     \\
$\Delta(1232)$  & 3/2 & $3m^b_q + 3a/(m^b_q)^2$       &   1239     \\
$\Lambda(1116)$ & 1/2 & $2m^b_q + m^b_s - 3a/(m^b_q)^2$ &   1114     \\
$\Sigma(1193)$  & 1/2 & $2m^b_q + m^b_s + a/(m^b_q)^2-4a/m^b_q m^b_s$ & 1179 \\
$\Sigma(1385)$  & 3/2 & $2m^b_q + m^b_s + a/(m^b_q)^2+2a/m^b_q m^b_s$ & 1381 \\
$\Xi(1318)$     & 1/2 & $2m^b_s + m^b_q + a/(m^b_s)^2-4a/m^b_q m^b_s$ & 1327 \\
$\Xi(1530)$     & 3/2 & $2m^b_s + m^b_q + a/(m^b_s)^2+2a/m^b_q m^b_s$ & 1529 \\
$\Omega(1672)$  & 3/2 & $3m^b_s + 3a/(m^b_s)^2$       & 1682 \\ \hline \hline
\end{tabular}
\end{center}
\end{table}

In Table \ref{tab:lqb} we summarize that description.  For all masses we use
values quoted by the Particle Data Group \cite{pdg} unless otherwise noted.
Effective masses of quarks in baryons and mesons can and do differ from one
another \cite{Lipkin:1978}, so we shall use superscripts $b$ and $m$ to denote
the former and latter.  The parameters of this table then may be interpreted as
summarizing all interactions between $qq$, $qs$, and $ss$.  We shall assume
these same interactions occur also in a baryon containing one $c$ or $b$ quark.
The average magnitude of the errors in this description is about 5 MeV.  We
shall use a similar method \cite{DeRujula:1975ge,Anikeev:2001rk}, with
appropriate corrections, to calculate masses of states with one or two heavy
quarks.

\subsection{Mesons}

A similar approach describes ground-state mesons composed of $u$, $d$, and $s$
quarks, as shown in Table \ref{tab:lqm}.  As effective masses of quarks in
mesons and baryons differ from one another, the parameters in Table
\ref{tab:lqm} will not be directly related to those in Table \ref{tab:lqb}.  We
do not discuss $\eta$, $\eta'$, whose masses are strongly affected by
octet-singlet mixing.  Here the average magnitude of errors is about 6 MeV.


\begin{table}
\caption{Quark model description of ground-state mesons containing $u,d,s$.
Here we take $m^m_u = m^m_d \equiv m^m_q = 310$ MeV, $m^m_s = 483$ MeV,
$b/(m^m_q)^2 = 80$ MeV.
\label{tab:lqm}}
\begin{center}
\begin{tabular}{c c c c} \hline \hline
State (mass     & Spin &   Expression for mass   & Predicted  \\
in MeV)         &      & \cite{Gasiorowicz:1981jz} & mass (MeV) \\ \hline
$\pi(138)$      &  0   & $2 m^m_q - 6b/(m^m_q)^2$    & 140 \\
$\rho(775),\omega(782)$ & 1 & $2 m^m_q + 2b/(m^m_q)^2$ & 780 \\
$K(496)$        &  0   & $m^m_q + m^m_s - 6b/(m^m_q m^m_s)$ & 485 \\
$K^*(894)$      &  1   & $m^m_q + m^m_s + 2b/(m^m_q m^m_s)$ & 896 \\
$\phi(1019)$    &  1   & $2 m^m_s + 2b/(m^m_s)^2$ & 1032 \\ \hline \hline
\end{tabular}
\end{center}
\end{table}
The overprediction of the $\phi$ mass may indicate slightly stronger binding
between two strange quarks.  We should keep this possibility in mind when
discussing other states with two strange quarks, but these do not occur for
$\Xi_{(cc,bb,bc)}$.  Some hint of this effect is also present when comparing
the predicted $M(\Xi)$ and $M(\Omega)$ with experiment, though the predicted
$M(\Xi^*)$ comes within 1 MeV of the observed value.

\section{States with one charmed quark \label{sec:onec}}

\subsection{Mesons}

We discuss mesons first because the $c \bar s$ interaction in $D_s^{(*)}$
displays a significant binding effect.  This is then related using a simple
QCD argument to the $cs$ binding in baryons, which is important to keep in
mind when predicting $\Xi_c^{(\prime,*)}$ and $\Omega_c^{(*)}$ masses.

The model of Sec.\ \ref{sec:uds} predicts
\beq \label{eqn:D}
M(D(1867.2))  = m^m_q + m^m_c - 6b/(m^m_q m^m_c),~
M(D^*(2008.6)) = m^m_q + m^m_c + 2b/(m^m_q m^m_c)~.
\eeq
The new parameter in these expressions is $m^m_c$, which may be estimated using
\beq
m^m_c = [3 M(D^*) + M(D)]/4 - m^m_q = (1973.3 - 310)~{\rm MeV} = 1663.3
~{\rm MeV}~.
\eeq
Using this value and $b/(m^m_q)^2 = 80$ MeV one estimates the hyperfine
splitting between $D$ and $D^*$ to be $M(D^*) - M(D) = 8b/(m^m_q m^m_c)=119.3$
MeV, to be compared with the observed value of 141.4 MeV.  Thus there seems
to be a hyperfine enhancement between $c$ and $\bar q$ relative to $q$
and $\bar q$.  This difference does not seem to occur between $cq$ and $qq$
hyperfine interactions, however, as we shall see when discussing charmed
baryons.

Charmed-strange mesons display an effect of enhanced $c \bar s$ binding.
Anticipating this, we may write
\bea \label{eqn:Ds} \nonumber
M(D_s(1968.5))   & = & B(c \bar s) + m^m_s + m^m_c - 6b/(m^m_s m^m_c)~,\\
M(D_s^*(2112.3)) & = & B(c \bar s) + m^m_s + m^m_c + 2b/(m^m_s m^m_c)~,
\eea
allowing one to solve for the binding term
\beq
B(c \bar s) = [3 M(D_s^*) + M(D_s)]/4 - m^m_s - m^m_c = - 69.9~{\rm MeV}~.
\eeq
This quantity will be related to the binding between $c$ and $s$ quarks
when we discuss charmed-strange baryons.
This term represents the additional binding to $c$ of the heavier $\bar s$
quark in comparison with that of the $\bar u$ or $\bar d$, due to the
shorter Compton wavelength of the $\bar s$ which allows it to sit more
deeply in the interquark potential.

Comparing Eqs.\ (\ref{eqn:D}) and (\ref{eqn:Ds}), one would conclude
that
\beq \label{eqn:hfds}
M(D_s^*) - M(D_s) = (m^m_q/m^m_s)[M(D^*) - M(D)] = 90.6~{\rm MeV}~,
\eeq
a factor of 0.63 times the observed value of 143.8 MeV which is almost the same
as $M(D^*) - M(D)$.  The scaling of the wave function describing the $c \bar s$
or $c \bar q$ bound state in a confining potential accounts for this behavior
\cite{Rosner:1990xx}.  We shall estimate the $cs$ hyperfine interaction in 
baryons directly from the $\Omega_c^*- \Omega_c$ splitting, finding a similar
enhancement with respect to the nominal value implied by Table \ref{tab:lqb}.

\subsection{Baryons}

An approach to charmed baryon masses similar to that leading to the predictions
for $u,d,s$ baryons in Table \ref{tab:lqb} must take account of enhanced
$cs$ binding and an enhanced $cs$ hyperfine interaction.  The effect of $cs$
binding may be related to $c \bar s$ binding by means of a color-SU(3)
argument.  The interactions between two quarks in various color states
are summarized in Table \ref{tab:color}.  The quarks in a $c \bar s$ meson
are in a color singlet, while a $cs$ pair in a baryon is in a color
antitriplet.  The $cs$ interaction strength in a color triplet is half that
of $c \bar s$ in a color singlet, so we shall assume, for every $cs$ pair in
a charmed-strange baryon, that
\beq
B(cs) = B(c \bar s)/2 = - 35.0~{\rm MeV}~.
\eeq
As we shall see, this provides a contribution of reasonable magnitude.

\begin{table}
\caption{Relative attraction or repulsion $\langle T_1 \cdot T_2 \rangle$
of quarks $Q \bar Q$ or $QQ$ in various states.
\label{tab:color}}
\begin{center}
\begin{tabular}{c c c} \hline \hline
State & Color & $\langle T_1 \cdot T_2 \rangle$ \\ \hline
$Q \bar Q$ &   1   & $-4/3$ \upstrut \\ 
$Q \bar Q$ &   8   & $1/6$  \\
   $Q Q$   & $3^*$ & $-2/3$ \\
   $Q Q$   &   6   & $1/3$  \\ \hline \hline
\end{tabular}
\end{center}
\end{table}

The scaling of energy levels linearly with coupling strength is not an
automatic feature.  In a power-law central potential of the form $V(r)
= \lambda r^\nu$, spacings $\Delta E$ of energy levels depend on $\lambda$
via the relation \cite{Quigg:1977dd} $\Delta E \propto \lambda^{2/(2+\nu)}$.
Thus, in the Coulomb potential $(\nu = -1)$ the Rydberg scales as $\alpha^2$;
harmonic oscillator level spacings $(\nu = 2$) scale as the square root of
the force constant; and $\Delta E \propto \lambda$ for a logarithmic potential,
which has been shown to interpolate not only between charmonium and
bottomonium interactions \cite{Quigg:1977dd}, but also to apply approximately
to $s \bar s$ excitations \cite{Martin:1980rm}.

The hyperfine splitting between $\Omega^*_c$ and $\Omega_c$ would be given
by $6a/(m^b_s m^b_c)$, but we shall parametrize it independently by replacing
$a$ with $a_{cs}$.  Accounting for enhanced $cs$ binding and hyperfine
interaction, the predictions for baryon masses then may be summarized in
Table \ref{tab:cbar}.  Here we have used the experimental value of
$M(\Lambda_c)$ in Table \ref{tab:cbar} to estimate $m^b_c = M(\Lambda_c) -
2 m^b_q + 3a/(m^b_q)^2 = 1710.5$ MeV.


\begin{table}
\caption{Quark model description of ground-state baryons containing one charmed
quark.  Here we take $m^b_u = m^b_d \equiv m^b_q = 363$ MeV, $m^b_s = 538$ MeV,
$m^b_c = 1710.5$ MeV, and $a/(m^b_q)^2 = 50$ MeV.  The spin of the $qs$ pair is
taken to be zero in $\Xi_c$ and one in $\Xi'_c$.
\label{tab:cbar}}
\begin{center}
\begin{tabular}{c c c c} \hline \hline
State ($M$     & Spin &   Expression for mass   & Predicted  \\
in MeV)         &      &                         & $M$ (MeV) \\ \hline
$\Lambda_c(2286.5)$ & 1/2 & $2 m^b_q + m^b_c - 3a/(m^b_q)^2$ & Input \\
$\Sigma_c(2453.4)$ & 1/2 & $2 m^b_q + m^b_c+a/(m^b_q)^2-4a/(m^b_q m^b_c)$
 & 2444.0 \\
$\Sigma_c^*(2518.1)$ & 3/2 & $2 m^b_q+m^b_c+a/(m^b_q)^2+2a/(m^b_q m^b_c)$ 
 & 2507.7 \\
$\Xi_c(2469.3)$ & 1/2 & $B(cs) + m^b_q + m^b_s + m^b_c -3a/(m^b_q m^b_s)$
 & 2475.3 \\
$\Xi'_c(2575.8)$ & 1/2 & $B(cs) + m^b_q + m^b_s + m^b_c + a/(m^b_q m^b_s)$ & \\ 
 & & $- 2a/(m^b_q m^b_c) - 2 a_{cs}/(m^b_s m^b_c)$ & 2565.4 \\
$\Xi^*_c(2645.9)$ & 3/2 & $B(cs) + m^b_q + m^b_s + m^b_c+a/(m^b_q m^b_s)$ & \\
 & & $+ a/(m^b_q m^b_c) + a_{cs}/(m^b_s m^b_c)$ & 2632.6 \\
$\Omega_c(2695.2)$ & 1/2 & $2B(cs) + 2 m^b_s + m^b_c + a/(m^b_s)^2
 - 4 a_{cs}/(m^b_s m^b_c)$ & 2692.1$^a$ \\
$\Omega_c^*(2765.9)$ & 3/2 & $2B(cs) + 2 m^b_s + m^b_c + a/(m^b_s)^2
 + 2 a_{cs}/(m^b_s m^b_c)$ & 2762.8$^a$ \\ \hline  \hline
\end{tabular}
\end{center}
\leftline{$^a$ Difference between experimental values used to
determine $6 a_{cs}/(m^b_s m^b_c) = 70.7$ MeV.}
\end{table}

The hyperfine splitting between $\Sigma_c^*$ and $\Sigma_c$ is predicted
to be $6a/(m^b_q m^b_c) = 63.7$ MeV, to be compared with the observed
value of 64.5 MeV.  Thus there does not seem to be an enhancement of
the hyperfine interaction between $c$ and $q$ over the value inferred
from Table \ref{tab:lqb}.

The states $\Xi_c$ and $\Xi'_c$ will mix with one another as a result of
SU(3) breaking.  This effect, leading to mass shifts of the order of several
MeV \cite{Rosner:1981yh}, has been ignored.

The naive hyperfine term $6a/(m^b_s m^b_c) = 43.0$ MeV is 0.61 times a
term $6 a_{cs}/(m^b_s m^b_c) = 70.7$ MeV evaluated using the splitting
between $\Omega^*_c$ and $\Omega_c$.  Thus the $cs$ hyperfine interaction
in baryons undergoes the same enhancement with regard to the naive value as
does the $c \bar s$ hyperfine interaction in mesons.

The average magnitude of the errors in the predictions of Table \ref{tab:cbar}
is about 9 MeV, not much higher than that for the light-quark baryons in
Table \ref{tab:lqb}.

\section{States with one $b$ quark \label{sec:oneb}}

\subsection{Mesons}

We discuss $B_s$ and $B^*_s$ mesons in order to estimate binding effects of a
$b$ quark with an $\bar s$ antiquark, so as to assess $bs$ binding in a
baryon, and in order to obtain an effective mass of a $b$ quark in a meson.
The model of Sec.\ \ref{sec:uds} predicts
\beq \label{eqn:B}
M(B(5279.4))  = m^m_q + m^m_b - 6b/(m^m_q m^m_b)~,~
M(B^*(5325.2)) = m^m_q + m^m_b + 2b/(m^m_q m^m_b)~.
\eeq
By a calculation similar to that in Sec.\ \ref{sec:onec}, one finds
\beq
m^m_b = [3M(B^*) + M(B)]/4 - m^m_q=(5313.8-363)~{\rm MeV} = 5003.8~{\rm MeV}~.
\eeq
The predicted hyperfine splitting is $M(B^*) - M(B) = 39.7$ MeV, a factor
of 0.87 times the observed value of 45.8 MeV.  For comparison, the predicted
hyperfine splitting $M(D^*)-M(D)$ was found in the previous Section to be
119.3 MeV, a factor of 0.84 times the observed value of 141.4 MeV.  This
near-equality is a consequence of the often-quoted relation
\beq
(45.78 \pm 0.35)~{\rm MeV} = M(B^*) - M(B) = (m^m_c/m^m_b) [M(D^*) - M(D)]
 = (47.0 \pm 0.1)~{\rm MeV}~,
\eeq
in which light-quark masses do not appear.

Allowing for a binding term $B(b \bar s)$, the pseudoscalar and vector $b
\bar s$ states have masses
\bea \nonumber
M(B_s(5366.77 \pm 0.24)) & = & B(b \bar s) + m^m_s + m^m_b -6b/(m^m_s m^m_b)~,\\
M(B^*_s(5415.4^{+2.4}_{-2.1})) & = & B(b \bar s) + m^m_s + m^m_b
 + 2 b/(m^m_s m^m_b)~,
\eea
where we have indicated errors on masses in MeV because those of $B^*_s$ are
non-negligible.  Repeating the calculation of the previous section, we find
\beq
B(b \bar s) = [3M(B^*_s)+M(B_s)]/4 - m^m_b - m^m_s
 = (-83.6 \pm 1.8)~{\rm MeV}~.
\eeq
This binding term is slightly larger than the value of $B(c \bar s)$ found
above, because the reduced mass of the $b \bar s$ system is greater than
that of $c \bar s$, leading to a shorter Compton wavelength and a more
deeply bound system.

The predicted hyperfine splitting between $B_s$ and $B^*_s$ is $8a/(m_b m_s) =
25.5$ MeV, to be compared with the observed value of $48.7^{+2.3}_{-2.1}$ MeV.
Alternatively, one may evaluate this quantity to be $m^m_q/m^m_s$ times the
observed value of $M(B^*) - M(B) = 45.8$ MeV, giving 29.4 MeV or a factor
of $0.60 \pm 0.03$ times the observed value.  For comparison, the same
scaling argument applied in Sec.\ \ref{sec:onec} gave $M(D_s^*) - M(D_s)$
a factor of 0.63 times its observed value.  Thus the relation
\beq
48.7^{+2.3}_{-2.1}~{\rm MeV} = M(B_s^*) - M(B_s) \simeq (m^m_c/m^m_b)
[M(D_s^*) - M(D_s)] = 47.8~{\rm MeV}~,
\eeq
in which light-quark masses do not appear, holds quite well.

\subsection{Baryons}

Recent progress in $b$-flavored baryon studies has been so great that we
have found it necessary to construct our own averages of masses.  These
are summarized in Table \ref{tab:bbarav}.  We have omitted measurements
superseded by those of higher statistics by the same collaboration, and
measurements older than 2011. 

We start with a value of the $b$ quark mass in baryons obtained from
the observed value of $M(\Lambda_b) = 5619.5 \pm 0.3$ MeV:
\beq
m^b_b = M(\Lambda_b) - 2 m_q + 3a/(m^b_q)^2 = 5043.5~{\rm MeV}
\eeq
The observed and calculated masses of the ground state $b$-flavored baryons are
summarized in Table \ref{tab:bbar}.  We note several points.

\begin{itemize}

\item Although the predicted $\Sigma_b$ and $\Sigma^*_b$ masses are a bit
below the observed ones, their predicted hyperfine splitting is 21.6 MeV,
while the observed value is $19.6 \pm 0.7$ MeV (neglecting a common systematic
error of 1.7 MeV).  Thus there is no evidence for enhancement of the term
$a/(m^b_q m^b_b)$ beyond the value based on Table \ref{tab:lqb}.

\item The rescaling of $a/(m^b_s m^b_b)$ to $a_{bs}/(m^b_s m^b_b)$ is
taken to be identical to that for the $c s$ hyperfine interaction in baryons,
which we saw was very close to that for the $c \bar s$ and $b \bar s$ mesons.
It could be tested in principle using the hyperfine difference prediction
\beq
M(\Omega_b^*) - M(\Omega_b) = 6 a_{bs}/(m^b_s m^b_b) = 24.3~{\rm MeV}~,
\eeq
but this involves detection of the very soft photon in the decay $\Omega_b^*
\to \gamma \Omega_b$, probably impossible.
The enhancement of $a_{cs}$ and $a_{bs}$ with respect to $a$ is due to the
deeper binding of the $c s$ and $b s$ system in comparison with $c q$
or $b q$, but a quantitative relation between $B(cs)$ and $a_{cs}$ or
between $B(bs)$ and $a_{bs}$ does not seem obvious to us.
A possible reason for lack of such a relation is that
$B(cs)$ and $B(bs)$ parametrize spin-independent binding,
while $a_{cs}$ and $a_{bs}$ measure the strength of a spin-dependent
interaction between the relevant quarks.

\item The predictions for $M(\Xi_b)$ and $M(\Omega_b)$ are not far from
those of Ref.\ \cite{Karliner:2008sv}:  $5795 \pm 5$ MeV and $6052.1 \pm 5.6$
MeV, respectively.  In that work some use was made of potential models,
whereas in the present estimates such effects are parametrized by binding
terms or modification of hyperfine interactions.

\item The average magnitude of errors in predictions of Table \ref{tab:bbar}
is about 8 MeV, a bit below that for charmed baryons in Table \ref{tab:cbar}.
We shall use these two errors and those in Table \ref{tab:lqb} to extrapolate
to the case of two heavy quarks, estimating prediction errors of 12 MeV
for $M(\Xi_{cc})$ and $M(\Xi_{bb})$.  For $M(\Xi_{bc})$ an additional
systematic error is associated with ignorance of the $B_c$--$B^*_c$ splitting.

\end{itemize}


\begin{table}
\caption{Averages of $b$-baryon masses based on recent experiments.
\label{tab:bbarav}}
\begin{center}
\begin{tabular}{l c c} \hline \hline
Baryon       &       Reference         & Mass (MeV) \\ \hline
$\Lambda_b$  & \cite{Aaij:2014pha}     & $5619.30 \pm 0.34$ \\
             & \cite{Aaltonen:2014wfa} & $5620.15\pm 0.31 \pm 0.47$ \\
             & \cite{Aad:2012bpa}      & $5619.7 \pm 0.7 \pm 1.1$ \\
             & Average                 & $5619.5 \pm 0.3$ \\
$\Sigma^+_b$ & \cite{Aaltonen:2011ac}  & $5811.3^{+0.9}_{-0.8}\pm1.7$ \\
$\Sigma^-_b$ & \cite{Aaltonen:2011ac}  & $5815.5^{+0.6}_{-0.5}\pm1.7$ \\
~~~Average$^a$ & (Over charges)        & $5814.26 \pm 1.76$ \\
$\Sigma^{*+}$ & \cite{Aaltonen:2011ac} & $5832.1 \pm 0.7^{+1.7}_{-1.8}$ \\ 
$\Sigma^{*-}$ & \cite{Aaltonen:2011ac} & $5835.1 \pm 0.6^{+1.7}_{-1.8}$ \\
~~~Average$^a$ & (Over charges)        & $5833.83 \pm 1.81$ \\
$\Xi_b^0$ & \cite{Aaij:2013pka}        & $5793.5 \pm 2.3$ \\
          & \cite{Aaltonen:2014wfa}    & $5788.7 \pm 4.3 \pm 1.4$ \\
          & \cite{Aaij:2014esa}        & $5791.80\pm0.39\pm0.17\pm0.26$ \\
          & Average                    & $5791.84 \pm 0.50$ \\
$\Xi_b^-$ & \cite{Aaij:2013qja}        & $5795.8 \pm 0.9 \pm 0.4$ \\
          & \cite{Aaltonen:2014wfa}    & $5793.4 \pm 1.8 \pm 0.7$ \\
          & Average                    & $5795.30 \pm 0.88$ \\
~~~Average & (Over charges)            & $5792.68 \pm 0.43$ \\
$\Xi_b^{*0}$ & \cite{Chatrchyan:2012ni} & $5949.71 \pm 1.25^{b}$ \\
$\Omega^-_b$ & \cite{Aaij:2013qja}     & $6046.0 \pm 2.2 \pm 0.5$ \\
             & \cite{Aaltonen:2014wfa} & $6047.5 \pm 3.8 \pm 0.6$ \\
             & Average                 & $6046.38 \pm 1.95$ \\ \hline \hline
\end{tabular}
\end{center}
\leftline{$^a$ Common systematic error added in quadrature.}
\leftline{$^b$ Ref.\ \cite{Chatrchyan:2012ni} quotes $M(\Xi_b^{*0}) -
M(\Lambda_b) - M(\pi^+) = (14.84 \pm 0.74 \pm 0.28)$ MeV.}
\end{table}


\begin{table}
\caption{Quark model description of ground-state baryons containing one bottom
quark.  Here we take $m^b_u = m^b_d \equiv m^b_q = 363$ MeV, $m^b_s = 538$ MeV,
$m_b = 5043.5$ MeV, and $a/(m^b_q)^2 = 50$ MeV.  The spin of the $qs$ pair is
taken to be zero in $\Xi_b$ and one in $\Xi'_b$.  The parameter $a_{bs}$
is rescaled from $a$ in the same manner as for charmed baryons:  $a_{bs}
= a_{cs} =(70.7/43.0)a$.
\label{tab:bbar}}
\begin{center}
\begin{tabular}{c c c c} \hline \hline
State ($M$     & Spin &   Expression for mass   & Predicted  \\
in MeV)         &      &                         & $M$ (MeV) \\ \hline
$\Lambda_b(5619.5)$ & 1/2 &  $2 m^b_q + m^b_b - 3a/(m^b_q)^2$ & Input \\
$\Sigma_b(5814.3)$ & 1/2 & $2 m^b_q + m^b_b+a/(m^b_q)^2-4a/(m^b_q m^b_b)$
 & 5805.1 \\
$\Sigma_b^*(5833.8)$ & 3/2 & $2 m^b_q+m^b_b+a/(m^b_q)^2+2a/(m^b_q m^b_b)$
 & 5826.7 \\
$\Xi_b(5792.7)$ & 1/2 & $B(bs) + m^b_q + m^b_s + m^b_b -3a/(m^b_q m^b_s)$
 & 5801.5 \\
$\Xi'_b(-)$ & 1/2 & $B(bs) + m^b_q + m^b_s + m^b_b + a/(m^b_q m^b_s)$
 & \\
 & & $- 2a/(m^b_q m^b_b) - 2 a_{bs}/(m^b_s m^b_b)$ & 5921.3 \\
$\Xi^*_b(5949.7)$ & 3/2 & $B(bs) + m^b_q + m^b_s + m^b_b + a/(m^b_q m^b_s)$
 & \\
 & & $+ a/(m^b_q m^b_b) + a_{bs}/(m^b_s m^b_b)$ & 5944.1 \\
$\Omega_b(6046.4)$ & 1/2 & $2B(bs) + 2 m^b_s + m^b_b + a/(m^b_s)^2
 - 4 a_{bs}/(m^b_s m^b_b)$ & 6042.8 \\
$\Omega_b^*(-)$ & 3/2 & $2B(bs) + 2 m^b_s + m^b_b + a/(m^b_s)^2
 + 2 a_{bs}/(m^b_s m^b_b)$ & 6066.7 \\ \hline  \hline
\end{tabular}
\end{center}
\end{table}

\section{Calculation of $ccq$ mass \label{sec:ccq}}

The mass of the $ccq$ state may be regarded as the sum of the following
contributions:

\begin{itemize}

\item The masses of the two charmed quarks

\item Their binding energy in a color $3^*$ state

\item Their mutual hyperfine interaction

\item Their hyperfine interaction with the light quark $q$

\item The mass of the light quark $q$

\end{itemize}

When more than one heavy quark is present, one must take into account
the binding energy between them.  We do this by comparing the sum of the
charm quark masses in the $1S$ charmonium levels $\eta_c$ and $J/\psi$ with
their spin-weighted mass
\beq
\bar M(c \bar c:1S) \equiv [3 M(J/\psi) + M(\eta_c)]/4 = 3068.6~{\rm MeV}~.
\eeq
We estimated the effective charm quark mass in a meson to be $m^m_c = 1663.3$
MeV.  The binding energy in $1S$ charmonium is thus $[3068.6-2(1663.3)]$ MeV
= --258.0 MeV.  Using the color-SU(3) relations in Table \ref{tab:color}
we then estimate the $cc$ binding energy in a baryon to be --129.0 MeV.

The $cc$ hyperfine interaction $a_{cc}/{m_c}^2$ is estimated as follows.  The
$\bar c c$ hyperfine splitting in the meson sector is given by $M(J/\psi) -
M(\eta_c) = 113.2 \ \hbox{MeV} = 4 a_{\bar c c}/(m^m_c)^2$. Assuming that the
quark-quark interaction $a_{c c}$ is half of the quark-antiquark interaction
$a_{\bar c c}$, and neglecting the small difference between $m^m_c$ and
$m^b_c$, we have $a_{c c}/(m^b_c)^2 {=} 1/2\cdot [M(J/\psi){-}M(\eta_c)]/4
{=}14.2$ MeV \cite{mbcmmc}.

We may then summarize the contributions to $M(\Xi_{cc})$ in Table
\ref{tab:xicc}.  The third line gives the contribution of the hyperfine
interaction between the two charmed quarks, while the fourth gives their
total hyperfine interaction with the light quark $q$.  The predicted value
$M(\Xi_{cc}) = 3627 \pm 12 $ MeV lies among a number of other estimates
summarized in Table \ref{tab:cccomp}, but well above the values claimed for
$\Xi_{cc}^+$ and $\Xi_{cc}^{++}$ by the SELEX Collaboration.

The hyperfine splitting is given by $M(\Xi_{cc}^*)- M(\Xi_{cc}) = 6a/m_q m_c
= 63.7$ MeV, yielding $M(\Xi_{cc}^*)= 3690 \pm 12$ MeV.  This state lies too
close in mass to $\Xi_{cc}$ to decay to it by pion emission, so it must
decay radiatively.

\begin{table}
\caption{Contributions to the mass of the lightest doubly charmed baryon
$\Xi_{cc}$.
\label{tab:xicc}}
\begin{center}
\begin{tabular}{c r} \hline \hline
Contribution & Value (MeV) \\ \hline
$2m^b_c + m^b_q$ & 3783.9 \\
$cc$ binding & ${-}129.0$ \\
$a_{cc}/(m^b_c)^2$ & 14.2 \\
${-}4a/m^b_q m^b_c$ & ${-}42.4$ \\
Total & 3627$\phantom{.0}\pm 12$ \\ \hline \hline
\end{tabular}
\end{center}
\end{table}

\begin{table}
\caption{Comparison of predictions for $M(\Xi_{cc})$.  
\label{tab:cccomp}}
\begin{center}
\begin{tabular}{c | c | c} \hline \hline
Reference & Value (MeV) & Method \\ \hline
Present work & $3627 \pm 12$ \\
\cite{DeRujula:1975ge} & 3550--3760 & QCD-motivated quark model \\
\cite{bj:heavy_baryon_masses} & $3668\pm 62$ & QCD-motivated quark model \\
\cite{Anikeev:2001rk} & 3651 & QCD-motivated quark model \\
\cite{Fleck:1989mb} & 3613 & Potential and bag models\\
\cite{Richard:1994ae} & 3630 & Potential model \\
\cite{Korner:1994nh} & 3610 & Heavy quark effective theory \\
\cite{Roncaglia:1995az} & $3660 \pm70$ & Feynman-Hellmann + semi-empirical \\
\cite{Lichtenberg:1995kg} & 3676 & Mass sum rules \\
\cite{Ebert:1996ec} & $3660 $ & Relativistic quasipotential quark model \\
\cite{SilvestreBrac:1996wp} & $3607 $ & Three-body Faddeev equations. \\
\cite{Gerasyuta:1999pc}&3527 & Bootstrap quark model + Faddeev eqs.\\
\cite{Itoh:2000um} & $ucc{:}\ 3649 \pm 12$, & \\
                   & $dcc{:}\ 3644 \pm 12$ & Quark model\\
\cite{Kiselev:2001fw} & $3480 \pm 50$ & Potential approach + QCD sum rules\\
\cite{Narodetskii:2002ib} & $3690 $ & Nonperturbative string \\
\cite{Ebert:2002ig} & $3620 $ & Relativistic quark-diquark\\
\cite{He:2004px} & $3520 $ & Bag model \\
\cite{Richard:2005jz} & 3643 & Potential model \\
\cite{Migura:2006ep} & 3642 & Relativistic quark model + Bethe-Salpeter\\
\cite{Albertus:2006ya} & $3612^{+17}$ & Variational \\
\cite{Roberts:2007ni} & $3678 $ & Quark model \\
\cite{Weng:2010rb} & $ 3540 \pm 20 $ & Instantaneous approx.\ +
Bethe-Salpeter\\
\cite{Zhang:2008rt} & $4260 \pm 190$ & QCD sum rules \\
\cite{Lewis:2001iz} & $3608(15)(^{13}_{35})$, & \\
                    & $3595(12)(^{21}_{22})$ & Quenched lattice\\
\cite{Flynn:2003vz} & 3549(13)(19)(92) & Quenched lattice\\
\cite{Liu:2009jc} & $3665 \pm 17 \pm 14^{+0}_{-78}$ 
 & Lattice, domain-wall + KS fermions\\
\cite{Namekawa:2012mp} & $3603(15)(16)$ & Lattice, $N_f=2+1$\\
\cite{Alexandrou:2012xk} & 3513(23)(14)&  LGT, twisted mass ferm.,
$m_{\pi}$=260 MeV \\
\cite{Briceno:2012wt} & 3595(39)(20)(6) & LGT, $N_f=2+1$, $m_{\pi}=200$ MeV\\
\cite{Alexandrou:2014sha} & 3568(14)(19)(1) 
& LGT, $N_f=2+1$, $m_{\pi}=210$ MeV \\ \hline \hline
\end{tabular}
\end{center}
\end{table}

\section{Calculation of $bbq$ mass \label{sec:bbq}}

One may apply very similar methods to calculate the mass of the lowest-lying
$\Xi_{bb}$ state.  The spin-weighted average of the $b \bar b:1S$ levels is 
\beq
\bar M(b \bar b: 1S) \equiv [3 M(\Upsilon) + M(\eta_b)]/4 = 9444.7~{\rm MeV}~.
\eeq
The spin-weighted average of the ground-state bottom mesons is
\beq
\bar M(b \bar q: 1S) \equiv [3 M(B^*) + M(B)]/4 = 5313.8~{\rm MeV}~.
\eeq
Subtracting $m^m_q = 310$ MeV, we arrive at $m^m_b = 5003.8$ MeV.  The binding
energy in $1S$ bottomonium is thus $[9444.7-2(5003.8)]$ MeV = --562.8 MeV.  By
arguments similar to those in the previous section, we then calculate the
binding energy between the two $b$ quarks in $\Xi_{bb}$ to be half this, or
--281.4 MeV.

\begin{table}
\caption{Contributions to the mass of the lightest baryon $\Xi_{bb}$ with
two bottom quarks.
\label{tab:xibb}}
\begin{center}
\begin{tabular}{c r} \hline \hline
Contribution & Value (MeV) \\ \hline
$2m^b_b + m^b_q$ & 10450.0\\
$bb$ binding & ${-}281.4$ \\
$a_{bb}/(m^b_b)^2$ & 7.8 \\
${-}4a/m^b_q m^b_b$ & ${-}14.4$ \\
Total & 10162$\phantom{.0} \pm 12$ \\ \hline \hline
\end{tabular}
\end{center}
\end{table}

\begin{table}
\caption{Comparison of predictions for $M(\Xi_{bb})$. 
\label{tab:bbcomp}}
\begin{center}
\begin{tabular}{c|c|c} \hline \hline
Reference & Value (MeV) & Method \\ \hline
Present work & $10162 \pm 12$ \\
\cite{bj:heavy_baryon_masses} & $10294\pm 131$ & QCD-motivated quark model \\
\cite{Anikeev:2001rk} & 10235 & QCD-motivated quark model \\
\cite{Richard:1994ae} & 10210 & Potential models\\
\cite{Roncaglia:1995az} & $10340 \pm100$ & Feynman-Hellmann + semi-empirical
formulas \\
\cite{Ebert:1996ec} & 10230 & Relativistic quasipotential quark model \\
\cite{Kiselev:2001fw} & $10090 \pm 50$ & Potential approach and QCD sum
rules \\
\cite{Narodetskii:2002ib} & $10160 $ & Nonperturbative string\\
\cite{He:2004px} & $10272 $ & Bag model\\
\cite{Roberts:2007ni} & $10322 $ & Quark model\\
\cite{Gerasyuta:2008zy} & 10045 & Coupled channel formalism \\
\cite{Weng:2010rb} & $ 10185 \pm 5 $ & Instantaneous approx.\ +
Bethe-Salpeter\\ 
\cite{Zhang:2008rt} & $9780 \pm 70$ & QCD sum rules \\
\hline \hline
\end{tabular}
\end{center}
\end{table}

The mass of a bottom quark in a baryon, m$^b_b = 5043.5$ MeV, was obtained in
Sec.\ \ref{sec:oneb}.  By the same approach as for $\Xi_{cc}$, the $bb$
hyperfine interaction term $a_{bb}/(m^b_b)^2$ may be taken as $(1/8)\cdot
[M(\Upsilon)-M(\eta_b)] = 7.8$ MeV \cite{mbcmmc}. 

We summarize the contributions to $M(\Xi_{bb})$ in Table \ref{tab:xibb}.  The
resulting value $M(\Xi_{bb}) =10162 \pm 12$ MeV tends to lie a bit below some
(but not all) estimates, as seen in Table \ref{tab:bbcomp}.

The hyperfine splitting is given by $M(\Xi_{bb}^*)-M(\Xi_{bb})=6a/m^b_q m^b_b =
21.6$ MeV, yielding $M(\Xi_{bb}^*)= 10184 \pm 12$ MeV.  This state decays
radiatively to $\Xi_{bb}$.

\section{Calculation of $bcq$ mass \label{sec:bcq}}

The methods of the previous two sections may be applied to calculate the
ground-state mass of $\Xi_{bc}$, with one qualification.  The $^3S_1$ state
of $b \bar c$, the $B_c^*$, has not yet been observed, so we shall have to
estimate its mass.  One method is to note that hyperfine interactions
between quarks with masses $m_1$ and $m_2$ are proportional to $|\Psi(0)|^2/
(m_1 m_2)$, so we need to evaluate the magnitude of $|\Psi(0)|^2$ for the
$b \bar c$ system by interpolating between $c \bar c$ and $b \bar b$.

A convenient parametrization is to assume that $|\Psi(0)|^2$ behaves as some
power $p$ of the reduced mass $\mu_R \equiv (m_1 m_2)/(m_1 + m_2)$.  With the
quark masses $m^m_c = 1663.3$ MeV and $m^m_b = 5003.8$ MeV and the hyperfine
splittings
\beq
M(J/\psi) - M(\eta_c) = 113.2~{\rm MeV}~,~~
M(\Upsilon) - M(\eta_b) = 62.3~{\rm MeV}~,
\eeq
one finds this power to be 1.46, very close to the value of 1.5 that one would
expect from a logarithmic potential.  Such a potential has been shown to
successfully interpolate between the charmonium and bottomonium spectra
\cite{Quigg:1977dd}, and now seems to give approximately the correct spacing
between the 1S and 2S of the $B_c$ system as well \cite{Aad:2014laa}.
With this power, the hyperfine splitting between $b$ and $\bar c$ in the
ground state is then estimated to be 68.0 MeV.  [This quantity also may be
estimated by taking the geometric mean of the charmonium and bottomonium
hyperfine splittings, with the result of 84.0 MeV.  The 16 MeV difference
between these two estimates can be viewed as an indication of the error
associated with determining $b\bar c$ hyperfine splitting].  The spin-weighted
average ground state $b \bar c$ mass is then
\beq
\bar M(b \bar c:1S) = M(B_c) + (3/4)(68.0~{\rm MeV}) = (6274.5 + 52.0)~
{\rm MeV} = 6325.5~{\rm MeV}~.
\eeq

The rest of the calculations proceed as in the previous two sections.  The
binding energy in the spin-weighted average $b \bar c$ ground state is
$6325.5 - 5003.8 - 1663.2 = {-}341.5$ MeV, so in a $bc$ baryon it is half this,
or --170.8 MeV.  The $bcq$ mass (before accounting for binding and hyperfine
interactions) is
\beq
m^b_b + m^b_c + m^b_q =
(5043.5 + 1710.5 + 363)~{\rm MeV} = 7117.0~{\rm MeV}~.
\eeq
The error associated with $c \bar s$ binding may be taken to be 3/4 times that
of the hyperfine splitting between $b$ and $\bar c$, or (3/4)(16 MeV) = 12 MeV.
We then take the error on the $cs$ binding to be 6 MeV.  The strength of the
$bc$ hyperfine interaction is determined by the same approach as for $\Xi_{cc}$
and $\Xi_{bb}$, i.e., $a_{bc}/(m^b_b m^b_c) = (1/8) \cdot b\bar c$\ hyperfine
splitting. As a result, a small error also is introduced to the $bc$ hyperfine
interaction.

The presence of three distinct quarks in $\Xi_{bc} = bcq$ means that there
are two ways of coupling them up to spin 1/2 in an S-wave ground state.
Taking the basis defined by the combined spin of the two lightest quarks,
as was done for the $\Xi_c = csq$ and $\Xi_b = bsq$, we call the state
with $S(cq) = 0$ the $\Xi_{bc}$ and that with $S(cq) = 1$ the $\Xi'_{bc}$.
Tables \ref{tab:xibc0} and \ref{tab:xibc1} show the respective contributions
to their masses, and Tables \ref{tab:bccomp0} and \ref{tab:bccomp1} compare our
predictions with others.  The $\Xi'_{bc}$ will decay radiatively to $\Xi_{bc}$.
The uncertainties on the masses of these two states are calculated by adding
in quadrature the spread between the two masses in each table and the global
error assumed to be 12 MeV.

The mass of the $J=3/2$ state is given by $M(\Xi_{bc}^*) = M(\Xi'_{bc}) +
3a/(m^b_q m^b_b) + 3a_{bc} /(m^b_b m^b_c) = M(\Xi'_{bc}) + 36.3$ MeV.  
Using the $M(\Xi'_{bc})$ value in the first column of
Table \ref{tab:xibc1} we then obtain $M(\Xi_{bc}^*)= 6969 \pm 14$ MeV.  As in
previous cases, this state decays radiatively to the $J=1/2$ ground state.

\begin{table}
\caption{Contributions to the mass of the lightest baryon $\Xi_{bc}$ with
one bottom and one charmed quark and the $cq$ pair in a spin-singlet state.
\label{tab:xibc0}}
\begin{center}
\begin{tabular}{c r r} \hline \hline
Contribution & Value (MeV) $\phantom{aa}$
& Value (MeV) $\phantom{aa}$ \\
& from$\phantom{aaaaa}$
& from$\phantom{aaaaa}$ \\
& $|\Psi(0)|^2 \sim \mu_{R}^{1.46}$ \mystrut
& \strut\kern2em $\sqrt{HF(\bar b b){\cdot}HF(\bar c c)}$
\\
\hline
$m^b_b + m^b_c + m^b_q$ & 7117.0 & 7117.0 \\
$bc$ binding & ${-}170.8$  & ${-}164.8$ \\
${-}3a/(m^b_c m^b_q)$ & --31.8 & --31.8 \\
Total & $6914\phantom{.0}\pm13$ & $6920 \phantom{.0}\pm13$ \\ \hline \hline
\end{tabular}
\end{center}
\end{table}

\begin{table}
\caption{Contributions to the mass of the lightest baryon $\Xi'_{bc}$ with
one bottom and one charmed quark and the $cq$ pair in a spin-triplet state.
\label{tab:xibc1}}
\begin{center}
\begin{tabular}{c r r} \hline \hline
Contribution & Value (MeV) $\phantom{aa}$
& Value (MeV) $\phantom{aa}$ \\
& from$\phantom{aaaaa}$
& from$\phantom{aaaaa}$ \\
& $|\Psi(0)|^2 \sim \mu_{R}^{1.46}$ 
& \strut\kern2em $\sqrt{HF(\bar b b){\cdot}HF(\bar c c)}$
\\
\hline
$m^b_b + m^b_c + m^b_q$ & 7117.0 & 7117.0 \\
$bc$ binding & ${-}170.8$  & ${-}164.8$ \\
$a/(m^b_c m^b_q)$ & 10.6 & 10.6 \\
$-2a/(m^b_b m^b_q) - 2a_{bc}/(m^b_b m^b_c)$ & --24.2 & --28.2 \\
Total & $6933\phantom{.0}\pm12$ & $6935 \phantom{.0}\pm12$ \\ \hline \hline
\end{tabular}
\end{center}
\end{table}

\begin{table}
\caption{Comparison of predictions for $M(\Xi_{bc})$. 
\label{tab:bccomp0}}
\begin{center}
\begin{tabular}{c| c| c} \hline \hline
Reference & Value (MeV) & Method \\ \hline
Present work & $6914 \pm 13$ \\
\cite{bj:heavy_baryon_masses} & $6916\pm 139$ & QCD-motivated quark model \\
\cite{Anikeev:2001rk} & 6938 & QCD-motivated quark model \\
\cite{Richard:1994ae} & 6930  & Potential models\\
\cite{Roncaglia:1995az} & $6990 \pm90$ & Feynman-Hellmann + semi-empirical
formulas \\
\cite{Lichtenberg:1995kg} & 7029 & Mass sum rules \\
\cite{Ebert:1996ec} & $ 6950 $ & Relativistic quasipotential quark model\\
\cite{SilvestreBrac:1996wp} & $6915 $ & Three-body Faddeev equations. \\
\cite{Kiselev:2001fw} & $6820 \pm 50$ & Potential approach and QCD sum
rules\\
\cite{Narodetskii:2002ib} & $6960 $ & Nonperturbative string \\
\cite{Ebert:2002ig} & 6933  & Relativistic quark-diquark\\
\cite{He:2004px} & $6800 $ & Bag model\\
\cite{Albertus:2006ya} & 6919  & Variational \\
\cite{Roberts:2007ni} & $7011 $ & Quark model\\
\cite{Gerasyuta:2008zy} & 6789 & Coupled channel formalism \\
\cite{Weng:2010rb} & $ 6840 \pm 10 $ & Instantaneous approx.\ +
Bethe-Salpeter\\ 
\cite{Zhang:2008rt} & $6750 \pm 50$ & QCD sum rules \\
\hline \hline
\end{tabular}
\end{center}
\end{table}

\begin{table}
\caption{Comparison of predictions for $M(\Xi_{bc}^\prime)$. 
\label{tab:bccomp1}}
\begin{center}
\begin{tabular}{c| c| c} \hline \hline
Reference & Value (MeV) & Method \\ \hline
Present work & $6933 \pm 12$ \\
\cite{bj:heavy_baryon_masses} & $6976\pm 99$ & QCD-motivated quark model \\
\cite{Anikeev:2001rk} & 6971 & QCD-motivated quark model \\
\cite{Roncaglia:1995az} & $7040\pm90$ & Feynman-Hellmann + semi-empirical
  formulas \\
\cite{Lichtenberg:1995kg} & 7053 & Mass sum rules \\
\cite{Ebert:1996ec} & $ 7000 $ & Relativistic quasipotential quark model\\
\cite{Kiselev:2001fw} & $6850 \pm 50$ & Potential approach and QCD sum
rules\\
\cite{Ebert:2002ig} & 6963  & Relativistic quark-diquark\\
\cite{He:2004px} & $6870 $ & Bag model\\
\cite{Albertus:2006ya} & 6948  & Variational \\
\cite{Roberts:2007ni} & $7047 $ & Quark model\\
\cite{Gerasyuta:2008zy} & 6818 & Coupled channel formalism \\
\cite{Zhang:2008rt} & $6950\pm80$ & QCD sum rules \\
\hline \hline
\end{tabular}
\end{center}
\end{table}

\section{P-wave excitations \label{sec:pw}}

In the event that a $\Xi_{(cc,bb,bc)}$ state is accompanied by a pion nearby
in phase space, the two can have come from a P-wave excitation.  Let us take
the example of $\Xi_{cc}$.

Heavy quark symmetry implies that in transitions involving a single pion
the $cc$ state maintains its spin of 1, while in such P-wave states the light
quark $q$ couples with a unit of orbital angular momentum to form a state of
total light-quark angular momentum $j=1/2$ or $j=3/2$.  We can then expect
a rich family of P-wave states with
\bea \nonumber
(j=1/2) \otimes (J(cc)=1) \to J_{\rm tot} & = & 1/2,~3/2~; \\
(j=3/2) \otimes (J(cc)=1) \to J_{\rm tot} & = & 1/2,~3/2,~5/2~.
\label{eqn:pw}
\eea
The parity of the $\Xi_{cc}$ is positive, whereas that of the states in
Eq.\ (\ref{eqn:pw}) is negative.  Heavy quark symmetry predicts that the
states with $j=1/2$ will decay via S wave pion emission, whereas states
with $j=3/2$ will decay via D wave pion emission, and hence will be narrower.
This is particularly true of the $J_{\rm tot} = 5/2$ state, which is pure
$j=3/2$ and hence immune from mixing.

Let us neglect the fine-structure interaction between the $j=3/2$
light-quark system and the heavy $cc$ diquark.  Even in P-wave
mesons with a single heavy quark, this interaction gives rise to
a splitting of only 41 MeV between $D_1(2421)$ and $D^*_2(2462)$,
and 20 MeV between $B_1(5723)$ and $B^*_2(5743)$.  The spin-weighted
average of $D_1(2421)$ and $D^*_2(2462)$ masses is 2446 MeV, lying
473 MeV above the spin-weighted average of $D$ and $D^*$ masses.
The spin-weighted average of $B_1(5723)$ and $B^*_2(5743)$ masses
is 5736 MeV, lying 422 MeV above the spin-weighted average of
$B$ and $B^*$ masses.  The $cc$ diquark is intermediate in mass
between the $c$ and $b$ quarks, so one might expect the narrow
P-wave excitations of $\Xi_{cc}$ to occupy an interval of no more
than a few tens of MeV, lying between 420 and 470 MeV above the
spin-weighted average of $\Xi_{cc}$ and $\Xi^*_{cc}$ masses.

\section{Likely decay modes and lifetimes \label{sec:br}}

Many of the references quoted in Tables \ref{tab:cccomp}, \ref{tab:bbcomp},
\ref{tab:bccomp0}, and \ref{tab:bccomp1} also discuss likely branching ratios
and production mechanisms.  In addition, we note early suggestions by Bjorken
\cite{bj1,bj2} and Moinester \cite{Moinester:1995fk}. Here we give some general
guidelines, avoiding specific calculations depending on details of form
factors and fragmentation.  We pay special attention to those modes which can
show up in the online selection criteria (``triggers'') of experiments at
$e^+ e^-$ colliders, the Tevatron, and the LHC.  We concentrate on those
decays involving the most-favored Cabibbo-Kobayashi-Maskawa matrix elements,
such as $c \to s W^{*+}$ and $b \to c W^{*-}$.  In lifetime estimates we
shall neglect the effects of Pauli interference, concentrating on effects of
factorized decays and $2 \to 2$ internal transitions.  Although we do not
present detailed branching fractions, Tables 9-18 through 9-20 of
Ref.\ \cite{Anikeev:2001rk} are a useful guide.

\subsection{$\Xi^{++}_{cc} = ccu$}

The decay of $\Xi^{++}_{cc}$ begins with the decay of either charm quark to a
strange quark and a virtual $W^+$ (``$W^{*+}$'').  In this and other processes,
a virtual $W^+$ gives rise to a positively charged hadronic state limited only
by available phase space.  In this case the minimum mass of the $csu$ remnant
is that of the $\Xi_c(2469)$.  Given our prediction of $M(\Xi_{cc}) = (3627 \pm
12)$ MeV, one has 1158 MeV of available energy for the $W^{*+}$ products, which
can then be $\pi^+$, $\rho^+$, or the low-energy tail of the $a_1^+$.

The $csu$ remnant has the quantum numbers of the $\Xi_c^+$.  It may decay via
virtual $W^+$ emission to an $ssu$ remnant which is either a $\Xi^0$ (hard to
detect) or an excited state of it (decaying to $\Xi^- \pi^+$).  Alternatively,
the $csu$ remnant may fragment into states such as $\Lambda^+_c K^- \pi^+$,
with the $\Lambda^+_c$ decaying to such final states as $p K^- \pi^+$.

The decay chain $\Xi_{cc}^{++} \to \pi^+ \Xi_c^+ \to 3 \pi^+ \Xi^-$ leads to
pions all of the same sign.  The CDF trigger based on two displaced tracks
accepts only a pair of opposite-sign tracks, and would miss such a signature
\cite{PTL}.  One might be able to pick up opposite-sign tracks from
higher-multiplicity decays giving rise to a $\pi^+$ and $\pi^-$ or $K^-$, but
one pays a price in higher multiplicity because such tracks are often soft and
below the accepted transverse momentum threshold.

A crude estimate of the lifetime of the $\Xi^{++}_{cc}$ may be obtained by
considering the two $c$ quarks to decay independently.
Bjorken \cite{bj1,bj2} and Fleck and Richard
\cite{Fleck:1989mb} estimate $\tau(\Xi^{++}_{cc}) \simeq 200$ fs by
this method.  We reproduce this value by assuming an initial state with
$M(\Xi_{cc}) = 3627$ MeV, a final state with $M(\Xi_c) = 2469$ MeV, a weak
current giving rise to $e \nu, \mu \nu,$ and three colors of $u \bar d$,
a kinematic suppression factor
\beq
F(x) = 1 - 8x + 8 x^3 - x^4 + 12 x^2 \ln(1/x)~,~~x_{cc} \equiv [M(\Xi_c)/
M(\Xi_{cc})]^2 = 0.4634~,
\eeq
and a factor of 2 to count each decaying $c$ quark.  The resulting decay rate is
\beq
\Gamma(\Xi^{++}_{cc}) = \frac{10~G_F^2 M(\Xi_{cc})^5}{192 \pi^3}F(x_{cc})
 = 3.56 \times 10^{-12}{\rm~GeV}
\eeq
leading to a predicted lifetime of $\tau(\Xi^{++}_{cc}) = 185$ fs.  In this
calculation two compensating effects have been neglected:  (i) a form
factor for the weak transition $\Xi_{cc} \to \Xi_c$, and (ii) the excitation
of $csu$ states above $\Xi_c^+$.  Here and elsewhere we have assumed
$V_{ud} = V_{cs} = 1$ for favored elements of the
Cabibbo-Kobayashi-Maskawa matrix.  A similar approach to semileptonic
decays of hadrons containing a single heavy quark has been shown to reproduce
observed rates with an accuracy of about 10\% \cite{Gronau:2010if}.

\subsection{$\Xi^{+}_{cc} = ccd$}

We treat this final state separately because, in addition to decaying via the
subprocess $c \to s u \bar d$ discussed in the previous subsection, it may
decay via the subprocess $c d \to s u$.  The decays of $\Lambda_c = cud$
($\tau = 200 \pm 6$ fs) and $\Xi_c^0 = csd$ ($\tau = 112^{+13}_{-10}$ fs) are
probably enhanced by this subprocess with respect to those of $\Xi_c^+=csu$
($\tau = 442 \pm 26$ fs), where it cannot occur.  By comparing the $\Xi_c^+$
and $\Xi_c^0$ decay rates, and including a factor of 2 for the two charmed
quarks participating in $cd \to su$, the enhancement to the decay rate becomes
$8.78 \times 10^{-12}$ GeV and the lifetime becomes $\tau(\Xi^+_{cc}) = 53$ fs.
Bjorken \cite{bj1,bj2} and Fleck and
Richard \cite{Fleck:1989mb} predict about 100 fs.

The subprocess $c d \to s u$ in $\Xi^{+}_{cc} = ccd$ leads to an excited $csu$
state without the $\pi^+$ emitted in $\Xi^{++}_{cc}$ decay.  The rest of the
discussion proceeds as for $\Xi^{++}_{cc}$, but with slightly more available
phase space.  In particular, the fragmentation of $csu$ into $\Lambda^+_c K^-
\pi^+$ gives rise to a slightly more energetic $K^-$, advantageous for the
CDF two-opposite-sign-track trigger.

\subsection{$\Xi^+_{bc} = bcu$}

A factorization approach similar to that described for the $\Xi_{cc}$ states
may be used to estimate one set of contributions to $\Xi_{bc} = bcq$ decays.
There are two contributing subprocesses:  $b \to c d \bar u$ and $c \to s u
\bar d$.  In the case of the first, the weak current can produce not only
$e \nu$, $\mu \nu$, and $\bar u d$, but also $\tau \nu$ and $\bar c s$.
An interesting consequence of the last is the decay $\Xi_{bc} \to J/\psi
\Xi_c$, allowed for both charge states of $\Xi_{bc}$.  The rate for this
decay should not exceed the total in which the weak current produces a $\bar c
s$ pair.  For the sake of a very crude estimate, we shall neglect the masses
of {\it all} allowed states produced by the weak current.

The $b \to c W^{*-}$ subprocess, under assumptions similar to those in the
previous subsections, gives rise to a partial decay rate
\beq
\Gamma(\Xi_{bc} \to W^{*-} \Xi_{cc}) = \frac{9~G_F^2 M(\Xi_{bc})^5}{192 \pi^3}
 F \{[M(\Xi_{cc})/M(\Xi_{bc})]^2\} |V_{cb}|^2 = 6.87 \times 10^{-13}~{\rm GeV},
\eeq
where we have used $|V_{cb}| = 0.04$ and have assumed massless final states of
$e \nu$, $\mu \nu$, $\tau \nu$, three colors of $\bar u d$, and three colors of
$\bar c s$. The $c\to s W^{*-}$ subprocess gives rise to a larger partial rate:
\beq
\Gamma(\Xi_{bc} \to W^{*-} \Xi_b) = \frac{5~G_F^2 M(\Xi_{bc})^5}{192 \pi^3}
 F \{[M(\Xi_b)/M(\Xi_{bc})]^2\} = 2.01 \times 10^{-12}~{\rm GeV}~.
\eeq
In principle for $\Xi^+_{bc} = bcu$ there should be a third contribution from
the subprocess $bu \to cd$.  However, the near-equality of the lifetimes
of $\Xi_b^0 = bsu$ and $\Xi_b^- = bsd$ \cite{Aaltonen:2014wfa,Stone:2014pra,%
RosnerCDFb}, as summarized in Table \ref{tab:Xibtau}, suggests that this
process carries little weight, so we shall neglect it.
The sum of the two contributions to the $\Xi^+_{bc}$ decay rate is then
$2.70 \times 10^{-12}$ GeV, yielding a lifetime of $\tau(\Xi^+_{bc}) = 244$ fs.

For the $b \to c W^{*-}$ subprocess, contributing to the decay of both
$\Xi_{bc}$ states, the virtual $W$ can easily produce a negative pion.
Subsequent decays of the $ccq$ intermediate state easily lead to a positive
pion, so the CDF trigger should be able to respond to a pair of opposite-sign
displaced tracks coming from $\Xi_{bc}$ decays.

\begin{table}
\caption{Lifetimes of $\Xi_b$ baryons (ps)
\label{tab:Xibtau}}
\begin{center}
\begin{tabular}{c c c c} \hline \hline
State & Ref.\ \cite{Stone:2014pra} & Ref.\ \cite{Aaltonen:2014wfa} &
Ref.\ \cite{RosnerCDFb}\\ \hline
$\Xi_b^-$ & $1.55^{+0.16}_{-0.09} \pm 0.03$ & $1.32 \pm 0.14 \pm 0.02$ &
  $1.36 \pm 0.15 \pm 0.02$ \\ 
$\Xi_b^0$ & $1.477\pm 0.026 \pm 0.014 \pm 0.013$ & & \\ \hline \hline
\end{tabular}
\end{center}
\end{table}

One effect which we have not considered is the internal $2 \to 2$ transition
$bc \to cs$.  For both $\Xi_{bc} = bcq$ states, this leads to a final
$csq$ state, an excited version of $\Xi^{(+,0)}_c$ which can decay to the same
products as $\Xi^{(+,0)}_c$ or hadronically to states like $\Lambda D^{0,+}$.
In principle one could relate the $bc \to cs$ process in $\Xi_{bc}$ to the
$b \bar c \to W^{*-}$ annihilation process in $B_c^-$ decay.
 
\subsection{$\Xi^0_{bc} = bcd$}

In addition to the contributions just calculated to the decay rate of
$\Xi^+_{bc}$, we have seen the subprocess $c d \to s u$ to be important
in the difference between $\Xi_c^0$ and $\Xi_c^+$ lifetimes. If we take
the additional contribution to the $\Xi_c^0$ decay rate to be the same
here, that provides an additional term of $4.39 \times 10^{-12}$ GeV,
leading to
\beq
\Gamma(\Xi^0_{bc}) = 7.09 \times 10^{-12}~{\rm GeV}~,~~
\tau(\Xi^0_{bc}) = 93~{\rm fs}~.
\eeq
The intermediate state produced by $c d \to s u$ is that of a excited
$bsu$ (``$\Xi_b^{*0}$'') with the mass of $\Xi_{bc}$.  The dominant subsequent
decay is governed by the subprocess $b \to c W^{*-}$, with enough phase space
that the virtual $W^-$ can produce all three lepton pairs, $\bar u d$, and
$\bar c s$.  The last process can lead to $J/\psi$ production, for example
in the decay $\Xi^0_{bc} \to J/\psi \Xi^0$ or $\Xi^0_{bc} \to J/\psi \Xi^-
\pi^+$.

\subsection{$\Xi_{bb} = bbq$}

Although the $2 \to 2$ process $b u \to c d$ is possible in principle for
$\Xi_{bb}^0 = bbu$, we have seen that it seems to play little role in
generating a lifetime difference between $\Xi_b^0$ and $\Xi_b^-$.  Hence we
may treat $\Xi_{bb}^0$ and $\Xi_{bb}^-$ generically as $\Xi_{bb} = bbq$ in
what follows.

The initial process in a $\Xi_{bb}$ decay is the process $bbq \to bcq +
W^{*-}$, where the minimum mass of the $bcq$ remnant is that of the
$\Xi_{bc}$, or 6914 MeV.  As the predicted mass of $\Xi_{bb}$ is 10162 MeV,
there is enough phase space for the weak current to produce all three lepton
pairs, $\bar u d$, and $\bar c s$.  Neglecting all of their masses, the
total decay rate is calculated to be
\beq
\Gamma(\Xi_{bb}) = \frac{18~G_F^2 M(\Xi_{bb})^5}{192 \pi^3}
F \{[M(\Xi_{bc})/M(\Xi_{bb})]^2\} |V_{cb}|^2 = 1.78 \times 10^{-12}~{\rm GeV}~,
\eeq
leading to a predicted lifetime $\tau(\Xi_{bb}) = 370$ fs.

An interesting decay involving the subprocess $b \to J/\psi~s$ {\it twice}
is the chain
\beq
\Xi_{bb} \to J/\psi~\Xi^{(*)}_b \to J/\psi~J/\psi~\Xi^{(*)}~,
\eeq
where $\Xi^{(*)}_b$ denotes a (possibly excited) state with the minimum mass
of $\Xi_b(5792)$, while $\Xi^{(*)}$ denotes a (possibly excited) state with the
minimum mass of $\Xi$.  Although this state is expected to be quite rare and
one has to pay the penalty of two $J/\psi$ leptonic branching fractions, it
has a distinctive signature and is worth looking for.

\subsection{Lifetime summary and discussion}

We summarize our lifetime predictions and compare them with others in
Table \ref{tab:life}.  There is quite a spread in predicted values, but
in all cases lifetimes are shortened when the $2 \to 2$ process $cd \to su$
is permitted, as in the case of the $\Lambda^+_c$, while the $2 \to 2$
process $bu \to cd$ seems to have little effect.  Our very short lifetime
for $\Xi^+_{cc}$ stems from two main effects:  (i) the difference between
the $\Xi_c^0$ and $\Xi_c^+$ lifetimes (112 vs 442 fs), used to estimate the
effect of the $cd \to su$ subprocess, and (ii) the factor
of 2 in the $cd \to su$ rate because the $\Xi^+_{cc}$ has two charmed quarks.


\begin{table}
\caption{Summary of lifetime predictions for baryons containing two heavy
quarks.  Values given are in fs.
\label{tab:life}}
\begin{center}
\begin{tabular}{c c c c c c} \hline \hline
Baryon & This work & \cite{Anikeev:2001rk} & \cite{Kiselev:2001fw} & \cite{bj2}
 & \cite{Moinester:1995fk} \\ \hline
$\Xi_{cc}^{++}=ccu$ & 185 & 430$\pm$100 & 460$\pm$50 & 500 & $\sim 200$ \\
$\Xi_{cc}^+=ccd$    &  53 & 120$\pm$100 & 160$\pm$50 & 150 & $\sim 100$ \\
$\Xi_{bc}^+=bcu$    & 244 & 330$\pm$80  & 300$\pm$30 & 200 & -- \\
$\Xi_{bc}^0=bcd$    &  93 & 280$\pm$70  & 270$\pm$30 & 150 & -- \\
$\Xi_{bb}^0=bbu$    & 370 &     --      & 790$\pm$20 & --  & -- \\
$\Xi_{bb}^-=bbd$    & 370 &     --      & 800$\pm$20 & --  & -- \\ \hline\hline
\end{tabular}
\end{center}
\end{table}

\section{Prospects for detection \label{sec:prod}}

Production of baryons containing two heavy quarks requires simultaneous
production of two heavy quark-antiquark pairs. Subsequently, a heavy quark
from one pair needs to coalesce with a heavy quark from the other pair, forming
together a color antitriplet heavy diquark. The heavy diquark then needs to
pick up a light quark to finally hadronize as a doubly-heavy baryon. The
coalescence of the two heavy quarks requires that they be in each other's
vicinity in both ordinary space and in rapidity space.  Computation of the
corresponding cross section from first principles is difficult
\cite{Anikeev:2001rk, Berezhnoy:1998aa,Savage:1990pr,SanchisLozano:1994vh,%
Doncheski:1995ye, Guberina:1999mx,Egolf:2002nk,Schmitt:2003gi,%
Koshkarev:2014ala,Kiselev:1994pu,Ma:2003zk}, and is subject to considerable
uncertainties due to nonperturbative effects.  Instead, we
use existing data \cite{Aaltonen:2007gv,Abazov:2008kv,Aaij:2012dd}
and theoretical estimates \cite{Brambilla:2004wf,Chang:2003cr,Gao:2010zzc} of
the closely-related process of $B_c$ production.

The two processes are closely related because production of $B_c$ also
requires simultaneous production of two heavy quark-antiquark pairs. A priori,
$B_c$ production has a somewhat higher probability, since in $B_c$ production a
heavy quark from one pair needs to coalesce with a heavy antiquark (rather than
a quark) from the other pair and there is no need to pick up an additional
light quark. There is no suppression associated with the latter, as once the
color anti-triplet heavy diquark is formed it can only hadronize by picking up
a light quark.  On the other hand, the attraction between a quark and an
antiquark is two times stronger than the attraction between two quarks and we
need to estimate the corresponding suppression factor.  In order to see if
$\Xi_{bc}$ and $B_c$ production rates are comparable, it would be useful to
compare the analogous production rates of $\Xi_c$ and $D_s$ (or $\Xi_b$ and
$B_s$) in experiments with large enough $E_{CM}$, whether in $e^+e^-$, $\bar p
p$, or $pp$ collisions.

Although it is not directly related, one may consider the relative probability
of a $b$ quark produced at high energy fragmenting into a meson (picking up
a light antiquark) and a baryon (picking up a light diquark).  The Heavy Flavor
Averaging Group (HFAG) \cite{Amhis:2012bh} has tabulated these quantities
as measured in Z decays and the Tevatron, as shown in Table \ref{tab:frag}.

\begin{table}
\caption{Fractions of different $b$-hadron species arising from $b$ 
quarks. From Ref.\ \cite{Amhis:2012bh}.
\label{tab:frag}}
\begin{center}
\begin{tabular}{c c c} \hline \hline
Quantity & $Z$ decays & Tevatron \\ \hline
$B^+$ or $B^0$ fraction $f_u = f_d$ & $0.403\pm0.009$ & $0.330\pm0.030$ \\
         $B^0_s$ fraction           & $0.103\pm0.009$ & $0.102\pm0.012$ \\
        $b$-baryon fraction         & $0.090\pm0.015$ & $0.236\pm0.067$ \\
\hline \hline
\end{tabular}
\end{center}
\end{table}

According to the HFAG analysis, depending on the production mechanism, the $b$
quark turns into a baryon between about 10 and 25\% of the time.  Fragmentation
into a baryon is somewhat favored at low transverse momentum
\cite{Amhis:2012bh} in hadron collisions.  

More recently, LHCb has carried out a thorough analysis of the 
$b$ quark fragmentation into mesons and baryons
\cite{Aaij:2011jp,Aaij:2013qqa,LHCb:2013lka,Aaij:2014jyk}.
In particular, the rather striking Fig.~4 in Ref.~\cite{Aaij:2014jyk}
shows that the ratio of $\Lambda_b$ production to $B^0$ meson production
for $p_T$ below 10 GeV is above 0.3 and goes above 0.5 for lower $p_T$.

A crude conclusion which we might draw from this comparison
is that a baryon composed of two heavy quarks could be produced with at
least as 10\% of the $B_c$ production rate.  An even more optimistic
estimate, supported by the above LHCb fragmentation data,
is provided by an explicit calculation \cite{Anikeev:2001rk} which predicts the
production rates for $\Xi_{cc}$ and $\Xi_{bc}$ to be as large as 50\% of that
for $(B_c+B^*_c)$ at the Tevatron, of the order of several nb.  The cross
section for $\Xi_{bb}$ is estimated in that work to be about a factor of 10
less.

The inclusive production cross section of the $B_c^+$ at the LHC, including the
contribution from excited states, was estimated to be $\sim 1\ \mu$b for
$\sqrt{s}=14$ TeV, and $\sim 0.4\ \mu$b for $\sqrt{s}=7$ TeV
\cite{Gao:2010zzc}, based on a dominant contribution
from $gg$ fusion: $gg\to B_c + b + \bar c$, computed by 
the complete order-$\alpha_s^4$ approach and by the fragmentation approach.

As a figure of merit, for 1 fb$^{-1}$ integrated luminosity $1\ \mu$b
translates to $\sim 10^9$ $B_c^+$ mesons being produced at the LHC, one order
of magnitude more than at the Tevatron. This number is considerably reduced
by triggering on specific decay modes and folding in the detector efficiency,
but nevertheless it leaves a sufficiently large number of $B_c$s to carry
out a detailed study of the $B_c^+$ properties.

Based on 0.37 fb$^{-1}$ of data collected in $pp$ collisions at $\sqrt{s}=7$
TeV LHCb has reported \cite{Aaij:2013voa} the ratio of the production
cross section times branching fraction between the $B_c^+ \to \jpsi \pi^+$
and the $B^+ \to \jpsi K^+$ decays,
\beq
\label{eq:Bc_over_Bplus_ratio_LHCb}
\frac{\sigma(pp{\to} B_c{+}X){\cdot}{\cal B}(B_c^+ {\rightarrow}\jpsi \pi^+)}
    {\sigma(pp{\to}B^+ {+}X){\cdot}{\cal B}(B^+{\rightarrow}\jpsi K^+)}
{=}
\ensuremath{(0.68 \pm 0.10\,({\rm stat.}) \pm
    0.03\,({\rm syst.}) \pm 0.05\,({\rm lifetime}) )}{\times}10^{-2},
\eeq
for $B_c^+$ and $B^+$ mesons with transverse momenta $p_{\rm T}>4$ GeV/$c$ and
pseudorapidities $2.5<\eta<4.5$, corresponding to $162 \pm 18$ $B_c^+ \to \jpsi
\pi^+$ signal events.  We may use this last figure to estimate the total number
of $B_c^+$ produced within the LHCb acceptance.

A number of calculations of $B_c$ branching fractions are compared with
one another in Ref.\ \cite{Ebert:2003cn}.  This reference is the one which
best reproduces the observed ratio \cite{Aaij:2014jxa}
\beq \label{eqn:pisl}
\frac{{\cal B}(B_c^+ \to J/\psi \pi^+)}{{\cal B}(B_c^+ \to J/\psi \mu^+ \nu)}
 = 0.0469 \pm 0.0028 \pm 0.0046~,
\eeq
so we shall quote its result ${\cal B}(B_c^+ \to J/\psi \mu^+ \nu) = 1.36\%$,
which we have corrected using a recent measurement \cite{Aaij:2014bva}
$\tau(B_c^+) = (509 \pm 8 \pm 12)$ fs.  With the measured ratio (\ref{eqn:pisl})
this implies ${\cal B}(B_c^+ \to J/\psi \pi^+) = 6.4 \times 10^{-4}$. 

With the above one can now compute the total $B_c$ production cross section 
directly from data:\footnote{We thank Vanya Belyaev for pointing out that
the total $B^+$ production cross section at LHCb is available and can be used
for this purpose.} the total $B^+$ production cross section at LHCb is
$38.9 \pm 0.3 (\hbox{stat.}) \pm 2.5 (\hbox{syst.}) \pm 1.3 (\hbox{norm.})$ 
$\mu$b \cite{Aaij:2013noa} 
and
${\cal B}(B^+ \to J/\psi K^+) = (1.028\pm0.031)\times 10^{{-}3}$ \cite{pdg}.
Putting this all together, we obtain
\bea
\label{eq:B_c_xsec_at_LHCb}
\sigma(pp{\to} B_c{+}X) &\approx&
\sigma(pp{\to}B^+ {+}X) \cdot
\frac{{\cal B}(B^+ \to \jpsi K^+)}
{{\cal B}(B_c^+ \to \jpsi \pi^+)}
\cdot 0.68\cdot 10^{{-}2}
\nonumber
\\
\\
&=&
{ 38.9 \cdot 1.028\times 10^{{-}3} \cdot 0.68\cdot 10^{{-}2}
\over
6.4 \times 10^{-4} 
}
\ \mu{\rm b}
= 0.4 \ \mu {\rm b}
\nonumber
\eea
for $4 < p_T < 40$ GeV and $2.5 < \eta < 4.5$, whereas Ref.\ \cite{Gao:2010zzc}
predicts this value for the whole of phase space.  With $162 \pm 18$ $B_c^+
\to\jpsi \pi^+$ events ${\cal B}(B_c^+ \to J/\psi \pi^+) = 6.4 \times 10^{-4}$ 
indicates a total of
\beq
\frac{162\pm18}{(6.4 \times 10^{-4})(0.0593 \pm 0.0006)} \sim 4.3 \times
10^6 ~B_c
\eeq
produced within the LHCb acceptance, where the second number in the denominator
is ${\cal B}(J/\psi \to \mu^+ \mu^-)$.  With an observed $B_c$
production cross section 0.4 $\mu$b in 0.37 fb$^{-1}$ there are a total of
about $1.5 \times 10^8$ $B_c$ produced overall, indicating an acceptance a
bit below 3\%.  One might expect the $\Xi_{cc}$ production cross section to
be at most a tenth of this, or 40 nb, at 7 TeV.

There is an interesting question whether $\Xi_{cc}$ is LHCb's best bet for
discovering doubly-heavy baryons.  The point is that because of Cabibbo
suppression the $b$ quark lifetime is about 7 times longer than the $c$ quark,
even though the $b$ quark is more than 3 times heavier and the phase space for
weak quark decay of a heavy quark scales like $(m_b/m_c)^5$ times a kinematic
function of the final and initial masses. Thus
$\tau(\Lambda_b) \approx 1.5\times 10^{-12}$ s
vs.\ $\tau(\Lambda_c) \approx 2\times 10^{-13}$ s, etc.  The difference between
actual $\Xi_{cc}$ and $\Xi_{bc}$ lifetimes, as shown in Table \ref{tab:life},
is not so pronounced.  Longer lifetime makes it much easier to identify the
secondary vertex.  On the other hand, the cross section for producing
bottom quarks is of course much smaller than for charmed quarks. So
there is a tradeoff.

For sake of completeness, we also provide here a brief update on the status of
search for doubly charmed baryons in $e^+e^-$ experiments.
The most recent and most stringent limits in this case come from Belle
\cite{Kato:2013ynr}. They used a 980 fb$^{-1}$ data sample to search for
$\Xi_{cc}^+$ and $\Xi_{cc}^{++}$ decaying into 
$\Lambda_{c}^{+}K^{-}\pi^{+}(\pi^{+})$
and $\Xi_{c}^{0}\pi^{+}(\pi^{+})$ final states.

Theoretical predictions for the inclusive cross section $\sigma(e^{+}e^{-}\to
\Xi_{cc} + X)$  at Belle CM energy, $\sqrt{s}=10.58$ GeV, vary over
a rather wide range, from 70 fb \cite{Kiselev:1994pu} to 230 fb
\cite{Ma:2003zk}.

Belle did not find any significant \,$\Xi_{cc}$\, signal and set a \,95$\%$ C.L.
upper limit on 
\break
$\sigma(e^{+}e^{-}\to \Xi_{cc}^{+(+)} + X)\times{\cal B}
(\Xi_{cc}^{+(+)} \to \Lambda_{c}^{+}K^{-}\pi^{+}(\pi^{+}))$
with the scaled momentum \hbox{$0.5{<}x_{p}{<}1.0$:}
4.1--25.0 fb for $\Xi_{cc}^{+}$ and 2.5--26.5 fb for $\Xi_{cc}^{++}$.
They also set a 95$\%$ C.L. upper limit on
$\sigma(e^{+}e^{-}\to \Xi_{cc}^{+(+)} + X) \times {\cal B} (\Xi_{cc}^{+(+)}
\to \Xi_{c}^{0}\pi^{+}(\pi^{+}))
\times{\cal B} (\Xi_{c}^{0} \to \Xi^{-}\pi^{+})$ with the scaled momentum
$0.45<x_{p}<1.0$:
0.076--0.35 fb for the $\Xi_{cc}^{+}$ and 0.082--0.40 fb for the
$\Xi_{cc}^{++}$.

The CM energy of the B factories is sufficient only for production of
$\Xi_{cc}$, as $\Xi_{bc}$ and $\Xi_{bb}$ are too heavy. So within the
foreseeable future the latter can only be produced at LHC and perhaps at
RHIC.

As in the case of doubly-heavy baryon production in LHCb, there is 
a significant uncertainty in theoretical predictions for the inclusive cross
section $\sigma(e^{+}e^{-}\to \Xi_{cc} + X)$.  Therefore, we suggest another
approach, similar in spirit to what we proposed for LHCb. This approach is
again directly based on observables which are in principle accessible in
$e^+e^-$ machines. 

One can make a rough estimate of the doubly-charmed baryon production rate by
assuming that the suppression of $ccq$ baryons $\Xi_{cc}$ vs.\ $csq$ baryons
$\Xi_c$ is of the same order of magnitude as the suppression of $\Xi_c$ vs.\
$ssq$ baryons $\Xi$. The physical content of this assumption is that the
suppression due to replacing an $s$ quark in a baryon by a much heavier $c$
quark is approximately independent of the spectator quarks in the baryon:
\beq
\label{ee_to_Xicc}
\sigma(e^{+}e^{-}\to \Xi_{cc} + X)
\sim
\sigma(e^{+}e^{-}\to \Xi_{c} + X)
\cdot
\frac{\sigma(e^{+}e^{-}\to \Xi_{c} + X)}
{\sigma(e^{+}e^{-}\to \Xi + X)}
\eeq

Information on inclusive $\Xi$ production in $e^+e^-$ annihilation 
at CM energy very close to Belle energy is readily available.
The ARGUS experiment has measured \cite{Albrecht:1987ew}
the following $\Xi^-$ rates per multihadronic event at $\sqrt{s}=10$ GeV: 
\bea
\label{ARGUS_Xi}
(2.06\pm 0.17\pm 0.23) \times 10^{-2} 
\qquad\hbox{in direct $\Upsilon$ decays}
\nonumber\\
\hbox{and}\
\phantom{aaaaaaaaaaaaaaaaaaaaaaaaaaaaaaaaaaaaaaaaaaaaaa}
\\
(0.67\pm 0.06\pm 0.07) \times 10^{-2} 
\qquad \hbox{in the continuum.}
\nonumber
\eea
The situation with inclusive $\Xi_c$ production is less simple.
Belle has seen $\Xi_c$ only in some specific channels, so what 
they measure is (production rate)$\times$(branching fractions into specific
channels). The latter are not known well, so it is not easy to
determine the production rate itself. 

Nevertheless, for our purpose it is sufficient to estimate the 
$\Xi_{cc}$ production rate to within a factor $2\div4$, which should be
possible even within the existing uncertainties about $\Xi_c$ branching
fractions.

The approximate formula in Eq.~(\ref{ee_to_Xicc}) and its generalizations to
$\Xi_{bc}$ and $\Xi_{bb}$ production should also apply to $pp$ collisions:
\bea
\label{pp_to_Xibc}
\sigma(pp \to \Xi_{bc} + X ) 
&\sim&
\sigma(pp \to \Xi_b + X)
\cdot
\frac{\sigma(pp \to \Xi_c + X)}
{\sigma(pp \to \Xi + X)}
\nonumber
\\
\\
&\sim&
\sigma(pp \to \Xi_c + X)
\cdot
\frac{\sigma(pp \to \Xi_b + X)} {\sigma(pp \to \Xi + X)}
\nonumber
\eea
as well as
\bea
\sigma(pp \to \Xi_{bb} + X ) 
&\sim&
\sigma(pp \to \Xi_b + X)
\cdot
\frac{\sigma(pp \to \Xi_b + X)} {\sigma(pp \to \Xi + X)}~.
\eea
\label{pp_to_Xibb}

\section{Conclusions} \label{sec:concl}

The conclusive observation of baryons with two heavy quarks is long overdue.
The weight of theoretical and experimental evidence suggests that whatever
the SELEX experiment has reported \cite{Mattson:2002vu,Ocherashvili:2004hi},
it is not the $\Xi_{cc}$:  Its mass lies below almost all expectations,
the isospin splitting between $\Xi^{++}_{cc}(3460)$ and $\Xi^+_{cc}(3520)$
candidates is implausibly large, and no other experiment has seen the effect.
We have predicted $M(\Xi_{cc}) = 3627 \pm 12$ MeV and made several suggestions
for its observation, including the decay to $\pi^+ \Xi_c$, where both states
of $\Xi_c^{+,0}$ have been identified in previous studies.  We also predict
the masses of other states summarized in Table \ref{tab:mpred}, and have
estimated lifetimes for these states as summarized in Table \ref{tab:life}.

\begin{table}
\caption{Summary of our mass predictions (in MeV) for lowest-lying baryons with
two heavy quarks.  States without a star have $J=1/2$; states with a star are
their $J=3/2$ hyperfine partners.  The quark $q$ can be either $u$ or $d$.  The
square or curved brackets around $cq$ denote coupling to spin 0 or 1.
\label{tab:mpred}}
\begin{center}
\begin{tabular}{c c c c} \hline \hline
State & Quark content & $M(J=1/2)$ & $M(J=3/2)$ \\ \hline
$\Xi^{(*)}_{cc}$ & $ccq$ &  $3627 \pm 12$ &  $3690 \pm 12$ \\
$\Xi^{(*)}_{bc}$ & $b[cq]$ & $6914 \pm 13$ &  $6969 \pm 14$ \\
$\Xi'_{bc}$      & $b(cq)$ & $6933 \pm 12$ &  -- \\
$\Xi^{(*)}_{bb}$ & $bbq$ & $10162 \pm 12$ & $10184 \pm 12$ \\ \hline \hline
\end{tabular}
\end{center}
\end{table}

We also estimate the hyperfine splitting between $B_c^*$ and $B_c$ mesons to be
$68$ MeV, with an alternate method giving 84 MeV.  P-wave excitations of the
$\Xi_{cc}$ with light-quark total angular angular momentum $j=3/2$, the analog
of those observed for $D$ and $B$ mesons, are estimated to lie around 420--470
MeV above the spin-weighted average of the $\Xi_{cc}$ and $\Xi^*_{cc}$ masses.
Production rates could be as large as 50\% of those for $B_c$, which
also requires the production of two heavy quark pairs.  We are optimistic that
with the increased data samples soon to be available in hadronic and $e^+ e^-$
collisions, the first baryons with two heavy quarks will finally be seen.

\section*{Acknowledgements}
The work of J.L.R. was supported by the U.S. Department of Energy, Division of
High Energy Physics, Grant No.\ DE-FG02-13ER41958.  We thank Vanya Belyaev,
Peter Cooper, Simon Eidelman, Aida El-Khadra, Lonya Frankfurt, Jibo He, Toru
Iijima, Patrick Lukens, Jean-Marc Richard, Sheldon Stone, Kalman Szabo, Mark
Strikman, and Jian-Rong Zhang for discussions of LHCb and Belle data and
relevant literature.  We also thank
James Bjorken for providing us with a copy of his unpublished manuscript
\cite{bj:heavy_baryon_masses} and Heath O'Connell of Fermilab Information
Resources for help in tracking it down.  M.K. would like to thank CERN Theory
Division for hospitality during the time when this work was finalized.

\section*{Note added}
After this work had been completed, a new set of lattice results appeared 
in Ref.~\cite{Brown:2014ena}. As noted by the authors, in several cases
their results are quite close to ours:
\hfill\break
$M(\Xi_{cc})  =   3610(23)(22)$ MeV,
$M(\Xi_{cc}^*)=   3692(28)(21)$ MeV,
$M(\Xi_{bb})  =  10143(30)(23)$ MeV,
$M(\Xi_{bb}^*)=  10178(30)(24)$ MeV,
$M(\Xi_{bc}) =    6943(33)(28)$ MeV,
$M(\Xi'_{bc}) =   6959(36)(28)$ MeV,
and
$M(\Xi_{bc}^*) =  6985(36)(28)$ MeV.


\begin{thebibliography}{99}
\bibitem{mbcmmc}
This approach ignores $m^b_c \neq m^m_c$ and $m^b_b \neq m^m_b$ .  Comparing
$a/(m^b_q)^2 = 50$ MeV with $b/(m^m_q)^2 = 80$ MeV, we have $a/b = 0.857$.  For
charm, the relevant comparison (since $a$ and $b$ already take into account the
factor of 2:  See Tables I and II) is $[a/(m^b_c)^2]/ [b/(m^m_c)^2] = 0.810$.
This would lower 14.2 MeV to 11.5 MeV.  Such an error added in quadrature would
not change our 12 MeV systematic error.  Similar (slightly smaller) shifts
are implied for the calculations of Tables IX and XI.
\end{thebibliography}
\end{document}